\documentclass[twocolumn]{aastex63}
\usepackage{gensymb}
\usepackage{graphicx}
\usepackage{amssymb}
\usepackage{amsmath}

\newcommand{\brg}{Br$\gamma$}
\newcommand{\ha}{H$\alpha$}
\newcommand{\kms}{\,km\,s$^{-1}$}
\newcommand{\mic}{\,$\mu$m}
\newcommand{\pds}{\object{PDS\,456}}
\newcommand{\qb}{\object{1H1613-097}}
\newcommand{\hi}{\ion{H}{1}}

\newcommand{\fua}{erg\,cm$^{-2}$\,s$^{-1}$\,\AA$^{-1}$}

\newcommand{\sbua}{erg\,cm$^{-2}$\,s$^{-1}$\,\AA$^{-1}$\,arcsec$^{-2}$}

\newcommand{\szero}{H$_2$ 1-0 S(0)}
\newcommand{\sone}{H$_2$ 1-0 S(1)}

\defcitealias{Fox_2015}{F15}
\defcitealias{Krishnarao_2020}{K20}
\defcitealias{Bordoloi_2017a}{B17}

\accepted{August 24, 2021}
\published{August 26, 2021}
\submitjournal{RNAAS} 

\shorttitle{A Search for Molecular Emission from the Fermi Bubbles}
\shortauthors{Fox, Kumari, Ashley et al.}
\graphicspath{{./}{figures/}}

\begin{document}

\title{A Near-Infrared Search for Molecular Gas in the Fermi Bubbles}
\correspondingauthor{Andrew Fox}
\email{afox@stsci.edu}

\author[0000-0003-0724-4115]{Andrew J. Fox}
\affil{AURA for ESA, Space Telescope Science Institute, 3700 San Martin Drive, Baltimore, MD 21218}

\author[0000-0002-5320-2568]{Nimisha Kumari}
\affil{AURA for ESA, Space Telescope Science Institute, 3700 San Martin Drive, Baltimore, MD 21218}

\author[0000-0002-6541-869X]{Trisha Ashley}
\affiliation{Space Telescope Science Institute, 3700 San Martin Drive, Baltimore, MD 21218}

\author[0000-0002-7705-2525]{Sara Cazzoli}
\affil{IAA - Instituto de Astrof{\'i}sica de Andaluc{\'i}a (CSIC), Apdo. 3004, E-18008 Granada, Spain}

\author[0000-0002-3120-7173]{Rongmon Bordoloi}
\affiliation{Department of Physics, North Carolina State University, 421 Riddick Hall, Raleigh, NC 27695-8202}

\begin{abstract}

We present Gemini/NIFS near-IR integral field spectroscopy of the 
fields-of-view around two AGNs
behind the Fermi Bubbles (\pds\ and \qb) to search for molecular gas in the 
Milky Way's nuclear wind. These two AGN sightlines were selected by the presence of 
high-velocity neutral and ionized gas seen in UV absorption. 
We do not detect any extended emission from the H$_2$ ro-vibrational 
S(0) and S(1) lines at 2.224 and 2.122\mic\ in either direction.
For the \sone\ line, the 3$\sigma$ surface brightness limits derived
from spectra extracted across the full $3''\times3''$ NIFS field-of-view are 
2.4$\times$10$^{-17}$\,\sbua\ for \pds\
and 4.9$\times$10$^{-18}$\,\sbua\ for \qb. Given these non-detections,
we conclude that CO emission studies and H$_2$ UV absorption studies 
are more promising approaches for characterizing molecular gas in the 
Fermi Bubbles.
\end{abstract}


\section{Introduction} \label{sec:intro}

Extending for 10\,kpc on either side of the Galactic Center,
the Fermi Bubbles trace an explosive event from the heart of the Milky Way \citep{Su_2010}.
The bubbles are filled with hot plasma at $T>10^6$\,K \citep{Crocker_2015} but
contain a population of cool embedded 
high-velocity clouds (HVCs).
These Fermi Bubble HVCs have been detected at many wavelengths, including
\hi\ 21\,cm emission \citep[][]{DiTeodoro_2018},
ultraviolet (UV) metal absorption 
\citep[][]{Fox_2015, Ashley_2020},
\ha\ emission \citep{Krishnarao_2020},
and sub-mm CO emission \citep{DiTeodoro_2020}.

In this research note we explore the Fermi Bubble HVCs in a new wavelength regime: 
near-IR spectroscopy. The near-IR contains the strong \szero\ and S(1) lines at 2.224\mic\ and 2.122\mic,
which trace warm molecular gas ($T\approx80$\,K), and \hi\ \brg\ at 2.166\mic, a tracer of 
warm ionized gas. 

\section{Observations and Data Reduction} \label{sec:redux}

\begin{figure*}[!ht]
\centering
\includegraphics[width=1.0\textwidth]{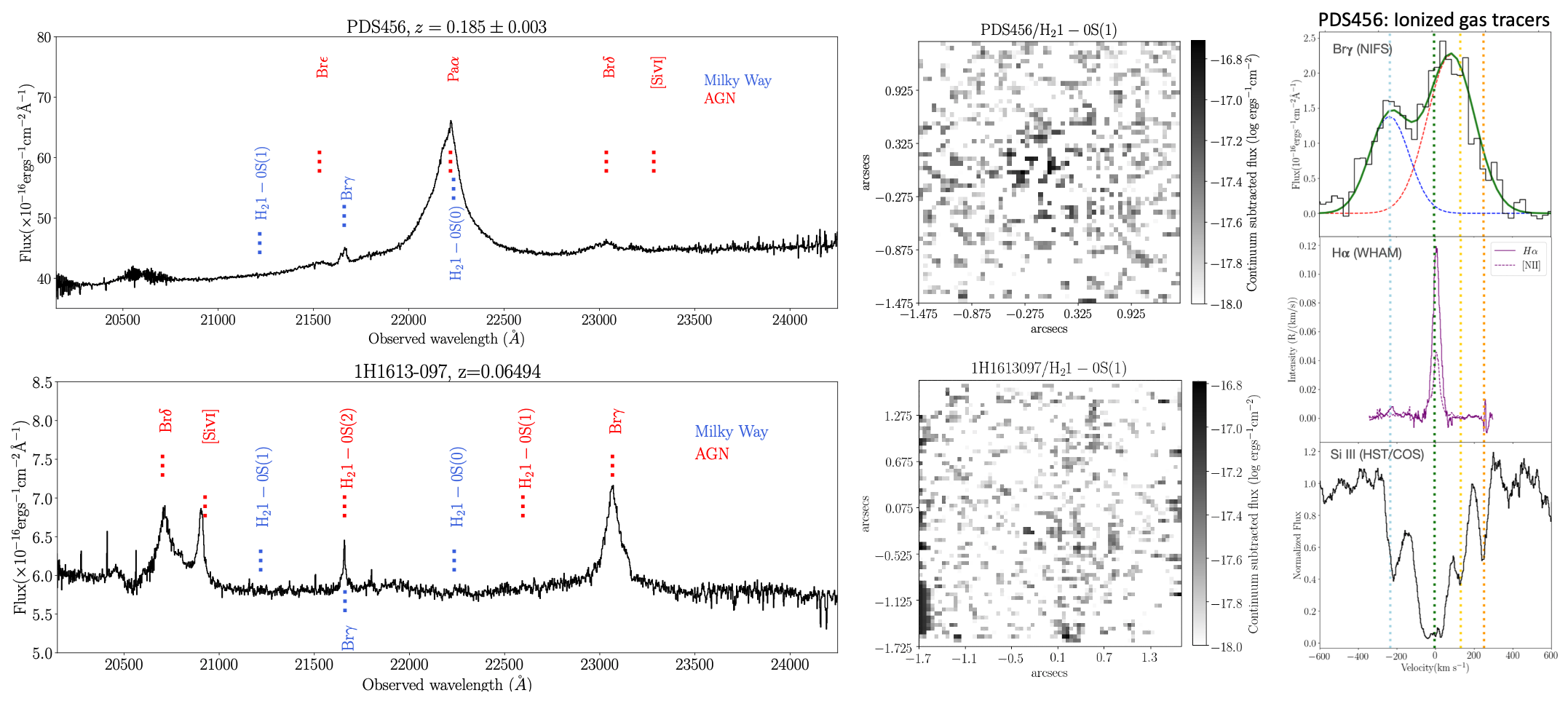} 
\caption{{\bf Left:} Gemini/NIFS K-band spectra for \pds\ (top) and \qb\ (bottom) 
extracted from the central circular regions of each field 0.35$''$ in radius. 
Galactic and AGN emission features are shown in blue and red, respectively.
{\bf Center:} Narrow-band continuum-subtracted images in the \sone\ line. 
The \szero\ lines also show no detection.
{\bf Right:} Comparison of line profiles toward \pds\ in three ionized-gas tracers: 
possible \brg\ emission (top; this work, see Section~\ref{sec:brackett}), 
\ha\ emission \citep[middle;][]{Krishnarao_2020}, 
and \ion{Si}{3} 1206 absorption \citep[bottom;][]{Fox_2015}. 
The dotted vertical lines indicate the \ion{Si}{3} component centroids.
}
\label{fig:montage}
\end{figure*}

We obtained Gemini-North Near-Infrared Integral Field Spectrometer (NIFS) 
observations of the \pds\ and \qb\ fields in April--May 2019 under Program ID 
GN-2019A-Q-323 (PI=A. Fox) using Natural Guide-Star mode. 

The targets \pds\ ($l,b,z_{\rm em}$=10.4$\degr$, 11.2$\degr$, 0.185) and \qb\ 
($l,b,z_{\rm em}$=3.4$\degr$, 28.5$\degr$, 0.06496) both lie behind the northern Fermi Bubble. They
were selected by their strong high-velocity UV absorption in 
\ion{Si}{3} $\lambda$1206, \ion{Si}{2} $\lambda$1260 and \ion{C}{2} $\lambda$1334
\citep[][]{Fox_2015, Bordoloi_2017a},
which ensures that high-velocity ionized and neutral gas is
present, increasing the chances for an IR emission-line detection.

The NIFS observations used the K-band grating covering the bandpass 1.99--2.44\mic. 
NIFS has a $3''\times3''$ field-of-view (FoV), a $0.1''\times0.04''$    
pixel size, and a K-band spectral resolution of $R$=5290, which allows for 
the separation of high-velocity GC emission ($|v|\gtrsim200$\kms) from foreground 
ISM emission ($v\sim0$\kms). 
We reduced the data using the NIFS Python data reduction 
pipeline \citep{Lemoine-Busserolle_2019}.
We determined the spatial offsets between different exposures for each 
target outside 
the pipeline and created a master data cube for each target by combining the 
spatially-aligned data cubes.

To conduct the sky subtraction, we used an $ABBA$ dither pattern with a separation
between 39--105$''$ for \pds\ and 53--98$''$ for \qb,
with the sky frames taken within 300--600\,s of the science frames.
We used 2MASS near-IR images \citep{Skrutskie_2006} to choose off-source 
directions with no detected emission. 

\section{Molecular Hydrogen} \label{sec:analysis}

To search for extended emission in the \szero\ and \sone\ 
emission lines, the bright central point source must be removed.
We extracted the AGN spectrum 
using a circular aperture of radius 0.35$''$,
chosen to match the median seeing of 0.7$''$. 
The resulting spectra are shown for both \pds\ and \qb\ in the left panels of Figure~\ref{fig:montage}.
The AGN spectra consist of power-law continua plus several discrete redshifted 
emission lines. We subtracted the continuum on a spaxel-by-spaxel basis 
by performing a linear fit to the 
AGN continuum around each line of interest. 
We created continuum-subtracted emission-line maps,
which are shown on the central panels of Figure~\ref{fig:montage}. 

For both the \pds\ and \qb\ directions, we find no detections 
of Galactic ($z$=0) emission in either \sone\ or S(0): the continuum-subtracted 
images show no signal in either direction.
We determined 3$\sigma$ upper limits on the 
flux density ($F$) and surface brightness density ($SB$) by measuring
the r.m.s. flux in the interval $-500\!<\!v\!<\!500$\kms\ in the continuum-subtracted 
spectra extracted over the full FoV ($3''\times3''$). 

Toward \pds, the derived limits are:\\
\\
$F_{\rm S(0)}\!<\!1.3\!\times\!10^{-16}$\,\fua\ ($<$2.2 mJy) and \\
$F_{\rm S(1)}\!<\!2.0\!\times\!10^{-16}$\,\fua\ ($<$3.1 mJy), i.e. \\ 
$SB_{\rm S(0)}\!<\!1.5\!\times\!10^{-17}$\,\sbua\ and\\
$SB_{\rm S(1)}\!<\!2.4\!\times\!10^{-17}$\,\sbua.\\
\\
Toward \qb, the limits are:\\
\\
$F_{\rm S(0)}\!<\!7.0\!\times\!10^{-17}$\,\fua\ ($<$1.2 mJy) and\\
$F_{\rm S(1)}\!<\!5.7\!\times\!10^{-17}$\,\fua, ($<$0.9 mJy), i.e.\\ 
$SB_{\rm S(0)}\!<\!5.9\!\times\!10^{-18}$\,\sbua\ and\\
$SB_{\rm S(1)}\!<\!4.9\!\times\!10^{-18}$\,\sbua.\\

The detection of CO emission in two Fermi Bubble clouds reported by 
\citet{DiTeodoro_2020} indicates that cold molecular gas is present in the nuclear wind
at $|b|<10\degr$.
Our near-IR non-detections of warm H$_2$ in the Fermi Bubbles 
at latitudes of 11.2$\degr$ and 28.5$\degr$
suggest a low covering fraction of warm molecular gas, 
especially given our pre-selection for 
high-velocity ionized and neutral gas, which should have maximized the chances of detecting molecular gas. 
We conclude that observations at other wavelengths, particularly
sub-mm CO emission-line studies and UV H$_2$ absorption-line studies, are more 
promising approaches for characterizing molecular gas in the Fermi Bubbles. 

\section{Emission at \brg} \label{sec:brackett}

Our NIFS data cubes cover Galactic \brg\ at 2.166 microns.
Toward \pds, a double-peaked emission component is seen at this wavelength 
(right panel of Figure~\ref{fig:montage}), 
with components near $-$235\kms\ and 0\kms.
This velocity structure closely mirrors the UV metal-line absorption profile 
\citep[][]{Fox_2015} and \ha\ emission profile
\citep[][]{Krishnarao_2020} in the \pds\ direction, suggesting the emission is a 
genuine detection of Galactic \brg. 
However, the emission is spatially 
unresolved and centered exactly on top of the QSO, 
with a radial profile that matches the continuum,
an unlikely arrangement if the emission traces a random foreground cloud.
The emission could be a redshifted QSO emission line;  
the closest candidate line 
is H$_2$ 1-0 S(5) at $z$=0.180, 
but this identification is uncertain owing to a $\approx$1500\kms\ offset 
from the CO emission redshift of $z_{\rm CO}$=0.1850$\pm$0.0001
\citep{Bischetti_2019}
and a lack of corresponding emission in the H$_2$ 1-0 S(3) line,
which is typically stronger. The origin of the emission is therefore uncertain.

Toward \qb, narrow emission at Galactic \brg\ is again detected, 
also unresolved and centered exactly on top of the QSO.
This emission could be due to H$_2$ 1-0 S(2) emission at $z$=0.06494, but there 
is no corresponding emission in H$_2$ 1-0 S(1), 
which should be stronger, so this identification is unconfirmed.

Given the contamination issues present in both 
sightlines, we do not present upper limits on the Galactic \brg\ emission.
Further \brg\ observations in unblended Fermi Bubble directions are needed to 
resolve the emission's origin.

\acknowledgments
We thank Julia Scharwaechter and Marie Lemoine-Busserolle 
for assistance with the Gemini/NIFS observations.
We thank Dhanesh Krishnarao, Howard Smith, Tracy Beck,
and Sylvain Veilleux for conversations.


\end{document}